# Subsurface Characteristics of Metal-Halide Perovskites Polished by Argon Ion Beam

Yu-Lin Hsu[1], Chongwen Li[2], Andrew C. Jones[3], Michael T. Pratt[3], Ashif Chowdhury[1], Yanfa Yan[2], and Heayoung P. Yoon*[1,4]

[1] Department of Electrical and Computer Engineering, University of Utah, Salt Lake City, UT 84112, USA

[2] Department of Physics and Astronomy and Wright Center for Photovoltaics Innovation and Commercialization, The University of Toledo, OH 43606, USA

[3] Center for Integrated Nanotechnologies, Materials Physics and Applications, Los Alamos National Laboratory, Los Alamos, NM 87545, USA

[4] Department of Materials Science and Engineering, University of Utah, Salt Lake City, UT 84112, USA

ABSTRACT

Focused ion beam (FIB) techniques have been frequently used to section metal-halide perovskites for microstructural investigations. However, the ion beams directly irradiated to the sample surface may alter the properties far different from pristine, potentially leading to modified deterioration mechanisms under aging stressors. Here, we combine complementary approaches to measure the subsurface characteristics of polished perovskite and identify the chemical species responsible for the measured properties. Analysis of the experimental results in conjunction with Monte Carlo simulations indicates that atomic displacements and local heating occur in the subsurface of methylammonium lead iodide (MAPbI$_3$) by glazing Ar$^+$ beam irradiation (≈ 15 nm by 4 kV at 3º). The lead-rich, iodine-deficient surface promotes rapid phase segregation under thermal aging conditions. On the other hand, despite the subsurface modification, our experiments confirm the rest of the MAPbI$_3$ bulk retains the material integrity. Our observation supports that polished perovskites could serve in studying the properties of bulk or buried junctions far away from the altered subsurface with care.

1. INTRODUCTION

Ion-beam based sample preparation techniques have been extensively used in studying the microstructural and interfacial characteristics of thin-film solar cells. Compared to traditional chemical and mechanical polishing, a focused ion beam (FIB) is fast and site-specific, minimizing mechanical damage, such as distorted structures, superficial scratching, and smearing at the surface [1-3]. For example, a liquid metal ion source (e.g., $^{31}$Ga) equipped with a typical dual-beam FIB/SEM (scanning electron microscopy) system produces a ray of ion beam at an accelerating voltage of 5 kV to 30 kV [4-5]. The focused beam (< 10 nm beam) irradiated on the target material introduces cascade events, resulting in the etching of the segment of interest. Post-processing methods are often applied to further remove or eliminate beam damage. The practices include low-energy FIB, gas-assisted etching, wet etching, and low-temperature milling [4, 6-7]. Previous efforts also demonstrated a broad argon ion beam ($^{18}$Ar; ≈ 1 mm beam) could effectively remove the FIB damage. Giannuzzi *et al.* showed a damaged region of less than 1 nm on a Si lamella after final thinning with 0.2 kV Ar$^+$ polishing [8].

Applications of FIB have enabled measuring morphology and local optoelectronic responses of metal-halide perovskite devices [9-10]. Smooth sample surfaces are desirable for many atomic/nanoscale measurements due to the scattering of the probe beams that can introduce artifacts arising from the sample topography rather than the characteristics of the sample of interest. Numerous transmission electron microscopy (TEM; beam energy of 80 keV to 300 keV) studies revealed the structural and compositional characteristics of perovskite solar cells (PSCs) under controlled environmental stressors [11-12]. High-fidelity Kelvin-probe force microscopy (KPFM) demonstrated the distinct potential distribution of cross-sectional PSCs under dark and light illumination, which can be correlated to ion migrations and defect dynamics [13-14]. While beneficial, recent publications have reported the beam sensitivity of hybrid organic-inorganic perovskites [10, 15-16]. For instance, PSC lamellas in TEM displayed significant deterioration during data acquisition via chemical-bond breakage and local heating. Cathodoluminescence (CL) spectra repeatedly collected on the same area of a FIB-milled MAPbI$_3$ (methylammonium lead iodide) presented the defect formation and irreversible decomposition within a few minutes [17]. Such rapid degradation seems mainly to be caused by high-energy electron beams. It is also possible the outer layer of the PSC is already altered during the FIB processing, accelerating fast degradation via modified mechanisms that may be far different from their pristine devices.

Herein, we thoroughly investigate the subsurface characteristics of perovskites exposed to an irradiating ion beam. We demonstrate a glazing angle Ar$^+$ milling to produce a smooth surface of MAPbI$_3$ (4 kV at 3º). The optoelectronic characteristics of the polished samples are measured by PL and KPFM and compared to their pristine controls. High-resolution X-ray Photoemission Spectroscopy (XPS) identifies the chemical species responsible for the observed PL and XPS, revealing a lead-rich and iodine-deficient surface induced by ion beams. Analysis of experimental results in conjunction with Monte-Carlo simulations indicates a thin layer of MAPbI$_3$ (≈ 15 nm) is damaged during the polishing while still retaining the material integrity in the rest of the film. We discuss the implications of our findings to study deterioration mechanisms of perovskites and possible uses of polished samples for surface insensitive measurements (e.g., two-photon microscopy studying buried junctions).

## 2. RESULTS AND DISCUSSION

Our typical perovskite solar cell consists of MAPbI$_3$ film sandwiched between an electron transfer layer (ETL) and a hole transfer layer (HTL). A fluorine-doped tin oxide (FTO) front contact on ETL and an Au back contact on HTL were used for photovoltaic measurements. In this study, we excluded the HTL and the Au contact to understand the direct impact of the Ar$^+$ beam on MAPbI$_3$. **Figure 1a** illustrates the configuration of the sample: Glass / FTO (200 nm) / SnO$_2$ (35 nm) / MAPbI$_3$ (550 nm). We applied Ar$^+$ beams to section the top surface of MAPbI$_3$. At an acceleration voltage of 4 kV, the broad beam (spot size > 1 mm) was irradiated at an incident angle of 3° parallel to the sample surface (**Figure 1b**). The polishing process was performed for 10 minutes at room temperature. By setting the beam focus to 0 %, the initial beam injected on the edge of the sample was widely spread on the sample surface in a trapezoidal shape. We characterized both polished and unpolished areas of the same sample along with an unexposed pristine perovskite sample. **Figures 1(c, d)** show representative SEM images of MAPbI$_3$ "before" and "after" the polishing. A typical grain size of the pristine sample ranged from 100 nm to 500 nm **(Figure 1c)**. As seen in **Figure 1d**, the glazing angle Ar$^+$ beam effectively removed the sample roughness but did not appreciably affect the film thickness. The porous structures near grain boundaries and the pinholes seen on the milled area are likely attributed to the grain growth during the thermal annealing after spin-coating of PSC precursors [18].

The primary focus in the present report is to analyze the subsurface characteristics rather than individual microstructural assessments. Luminescence characteristics were measured using PL spectroscopy at a laser beam excitation of 244 nm. At a nominal beam size of ≈ (400 μm)$^2$, the laser power was set to ≈ 10 μW to reduce possible light degradation during the measurements.

We used FDTD (finite-difference time-domain) simulations to obtain the absorption profile in MAPbI$_3$ under 244 nm illumination by using the refractive indices extracted from an ellipsometry study [19-21]. The simulation results suggest the illuminated photons ($\lambda$ = 244 nm) are mainly absorbed within the subsurface of < 100 nm, indicating the PL probe can be sensitive to subsurface modification. Details of the simulation can be found in the Supplementary Section.

**Figures 2(a, b)** display the representative PL spectra collected on the polished region compared to the unexposed area. The prominent peak at 765 nm (1.62 eV) in **Figure 2a** corresponds to the bandgap of MAPbI$_3$, showing an excellent agreement with the previous studies [22-23]. This bandgap emission decreases by 40 % after the milling, but the overall intensity is still strong. In recent literature, Kosasih *et al.* reported a bandgap emission on a FIB-prepared MAPbI$_3$ lamella (8 kV Ga$^+$ beam) using CL, where the peak near 1.6 eV was blue-shifted and broadened ($\approx$ 0.2 eV) compared to their bulk controls [17]. The authors suggested this behavior could be linked to the mechanical stress causing suppression of the atomic orbital overlap and an amorphization of MAPbI$_3$. In contrast, our PL peak position of 1.62 eV and the full-width at half-maximum (FWHM) of 0.09 eV remain unchanged after Ar$^+$ beam polishing. The weak peaks at 517 nm (2.4 eV) correspond to the PbI$_2$, indicating ion beams might degrade a small portion of the surface. The consistent PL characteristic was observed across the large area of the Ar$^+$ milled sample. **Figure 2c** shows a PL intensity map collected in an area of 8 mm × 2.6 mm at a step size of 800 μm. The color contrast reflects the peak height at 1.62 eV (765 nm). Uniform and strong PL responses are apparent in the unmilled regions (in red), whereas the Ar$^+$ polished region shows relatively lower PL intensities (in blue). **Figure 2d** plots the magnitudes of the PL height extracted from three lines marked in **Figure 2c**. The high luminescence obtained on the unexposed area (x < 1 mm, x > 6 mm; $\approx$ 5 × 10$^4$ counts per second [cps]) reduced to approximately 40 % in the center (1 mm < x < 6 mm; ≈ 3 × 10$^4$ cps), where the Ar ion beam was irradiated. The FWHM of all PL spectra ranges within a little variation (< 0.01 eV) and does not show any significant peak broadening across the sample. The overall PL intensity after the milling is still strong, implying the integrity of MAPbI3 bulk preserves.

We examine the electrostatics of the MAPbI$^3$ surface after Ar$^+$ milling using KPFM that measures the contact potential difference (CPD; $V_{CPD}$) between the probe tip and the sample surface.

$$V_{CPD} = \frac{\Phi_{tip} - \Phi_{sample}}{e} \tag{1}$$

Here $\Phi_{sample}$ is the work function of MAPbI3, $\Phi_{tip}$ is the work function of the probe tip, and e the elementary charge (1.6 × 10$^{-19}$ C) [24]. In our measurements, a probe tip (Pt/Ir-coated Si cantilever) was raster scanned on MAPbI3 at a scan rate of 0.5 Hz in a tapping mode while the FTO layer on the sample was grounded under an ambient environment. Topography and the CPD raw datasets were simultaneously recorded. The grain size of pristine MAPbI3 ranges from 100's nm to ≈ 1 μm, and the peak-to-valley roughness is approximately ≈ 100 nm (**Figure 3a**). As seen in **Figure 3b**, this surface roughness notably reduces to 10's nm after polishing it with an Ar$^+$ beam (4 kV irradiated at 3º for ten minutes). The topography of the milled sample presents porous structures and pinholes, primarily near grain boundaries. This porosity is likely attributed to the grain growth during the thermal annealing after spin-coating of PSC precursors [18]. **Figures 3 (c, d)** display the corresponding CPD distribution of the pristine and polished samples, showing relatively uniform surface potentials rather than distinct electronic structures at individual microstructures. A careful KPFM study by Lonzoni *et al.* found that an apparent CPD contrast of MAPbI3 microstructures became weaker under an ambient condition than in a vacuum when measuring the same sample [25]. Our low CPD contrast indicates possible partial oxidation at the surface. For a quantitative comparison, we determine the work function of MAPbI3 from equation (1). The mean $V_{CPD}$ value of the pristine and the polished samples is approximately -179 mV and 328 mV, respectively, calculated from the CPD maps in **Figures 3 (c, d)**. The work function of the probe tip, $\Phi_{tip} \approx 4.24$ eV $\pm 0.15$ eV, was first calibrated by measuring the CPD between the AFM probe and standard references of Au ($\Phi_{Au} \approx 4.6$ eV) film and HOPG (highly oriented pyrolytic graphite; $\Phi_{HOPG} \approx 5.1$ eV) [26]. Based upon the calibrated work function of the AFM probe, we calculate a work function of approximately 3.91 eV $\pm$ 0.15 eV for pristine MAPbI3 (Glass / FTO / SnO$_2$ / MAPbI3). This estimation reasonably agrees with the reported work function [27]. The notable CPD reduction after the polishing ($\Delta V_{CPD} \approx 510$ mV) suggests that the irradiating Ar$^+$ beam can further impact the electronic structures of MAPbI3 (sub)surface more than partial oxidation in air exposure.

To directly probe the elemental changes of MAPbI3 responsible for the observed PL and KPFM, we carry out X-ray photoemission spectroscopy (XPS). At a nominal beam spot size of 300 μm × 700 μm, XPS resolves the local chemical information near the surface (< 10 nm). The survey spectra acquired on a pristine and a polished sample are shown in **Figure S3**. Beginning with a pristine sample, the peaks of organic elements of C 1s, N 1s and O 1s are observed near 285 eV, 402 eV, and 533 eV, respectively. Four core level peaks of inorganic atoms are also apparent for this pristine sample: I 4d (40 eV ~ 60 eV), Pb 4f (130 eV ~ 150 eV), Pb 4d (410 eV ~ 450 eV), and I 3d (610 eV ~ 650 eV) [28-29]. After the ion beam exposure, the XPS peaks of the organic components are significantly reduced or almost diminished, while the inorganic atoms (Pb, I) showing strong peaks.

To gain better understanding, we perform in-depth analysis using high-resolution XPS profiles. **Figure 4a** compares the core-level O 1s spectra of the samples. While relatively weak, the O 1s signal of the pristine control indicates partial oxidation of the surface, evidence of the low CPD contrast of microstructures observed in KPFM (**Figure 3c**). The O 1s signature near 533 eV and the N 1s peak near 401.5 eV (**Figure 4b**) were almost vanished after Ar$^+$ milling. **Figure 4c** shows the core-level 1s spectra of C. This broad C 1s peak contains convoluted signatures of C-C, C-H, and C-N bonding [30]. The significantly reduced peak intensity indicates the ionized Ar atoms physically remove the lighter elements (i.e., C, N) at the surface. The dangling bonds of C 1s can be susceptible to forming COO- bonds in ambient air, accelerating the chemical deterioration of MAPbI$_3$.

On the other hand, the inorganic components (I, Pb) show noticeably enhanced XPS signals. The peaks of the doublet states of I 3d (619.5 eV and 631 eV; **Figure 4d**) and Pb 4f (138.5 eV, 143.5 eV; **Figure 4e**) substantially increase with Ar$^+$ milling. Interestingly, the Pb 4f spectrum of the polished sample shows the additional new peaks near the shoulders of the Pb 4f doublet peaks (blue dot lines in **Figure 4e**). We decompose the peaks using a Gaussian-Lorentz model, confirming the new peaks of metallic lead [Pb (0); 137.1 eV and 142 eV] and oxidized lead [Pb (II); 138.7 eV and 143.6 eV] [30-31]. As Kumar *et al.* suggested, the presence of the unsaturated Pb defects in the MAPbI$_3$ lattice can further oxidize it to lead monoxide (PbO) under thermal aging processes [30]. In **Figure 4f**, we show a summary of the compositional changes after Ar ion polishing. The overall I/Pb ratio decreases from 2.6 of pristine to 2.1 with polishing, implying the MAPbI$_3$ surface may be partially decomposed to lead iodide (PbI$_2$) or their complexes. Our XPS analysis suggest that the ion beam irradiation alters the atomic structure of the surface, and the locally damaged subsurface resulting from broken bonds can further affect phase segregation and degradation of MAPbI3 under aging processes.

We qualitatively examine how the modified MAPbI3 surface under ion beams influences bulk deterioration under a heat stressor. A separate sample with a geometry illustrated in **Figure 1** was prepared by $Ar^+$ milling (i.e., 4 kV $Ar^+$ beam injected at an angle of 3º parallel to the sample surface). By setting the beam focus of 0 %, the ion beam propagates on the sample surface in a trapezoidal shape (etched zone marked in a red dot line) while preserving unirradiated areas on the sample. A heat lamp facing the polished sample surface was placed above a sample stage, maintaining the sample temperature of 150 °C. **Figure 5** shows a series of sample images taken at different times. We observe insignificant color change for the first 10 minutes. At ≈ 20 minutes, visible yellow spots are seen near the edge of the ion-beam exposed area, which is continuously expanding to the entire polished region (trapezoidal zone) with time. An apparent color change of the unpolished zones is seen after ≈ 50 min heating. At ≈ 80 min, the color of the entire sample was turned from dark brown (MAPbI3) into vivid yellow (PbI2), indicating complete compositional dissociation. Our observation is consistent with the previous studies demonstrating an irreversible chemical dissociation of MAPbI3 into PbI2 at around 150 °C to 200 °C [32]. It is not surprising to observe the fast-thermal degradation on the ion-beam exposed area due to initial compositional alteration during the $Ar^+$ milling. However, we observe notably different degradation speeds, where the unexposed areas show marginal color changes (≈ 40 minutes) until the entire area exposed to the $Ar^+$ beam was segregated into PbI2. Previous studies suggest that the defect states at the surface and interfaces play a more significant role than point defects in bulk in their degradation mechanisms [33-34]. Our preliminary experiment supports the sensitivity of the surface and interface states of metal-halide perovskites in thermal aging processes.

To elucidate the detailed dynamics of the ion beam with perovskites, we simulate atomic displacements in MAPbI3 by Ar$^+$ beams using Stopping and Range of Ions in Matter (SRIM) and Transport of Ions in Matter (TRIM). This program package is built on the fundamental knowledge about atomic interaction, collisions, and transport theory, frequently used for calculating the damage profile and implanted ion range of various materials [35-36]. SRIM calculates the electronic stopping power by fitting experimental data, whereas TRIM simulates the energy loss of the incident atom to target electrons and the knock-on atoms along the ion path in the material. In our full cascade model, 200,000 Ar atoms at 4 kV were injected at an incident angle of 3° parallel to the surface of MAPbI3 (**Figure 1**). The standard data in the program set the material parameters of Ar and MAPbI3 (e.g., atomic weight, displacement energy, lattice energy, surface energy, etc.) except for the mass density of MAPbI3 (4.16 g/cm$^3$), which was obtained from the literature [37].

**Figure 6** displays the estimated damage distribution in MAPbI3 caused by the Ar ion beam. As seen in **Figure 6a**, the injected 4kV Ar ions penetrate MAPbI3 as high as 25 nm. The energy loss of the ions and their recoils occurs near the surface (5 nm), causing irreversible ionization. **Figure 6b** shows that part of the energy loss can contribute to the lattice vibration, primarily attributed to the recoils (**Figure 6b**). The calculated total energy loss is distributed into ionization of ≈ 47 %, vacancies of ≈ 6 %, and phonons of ≈ 47 %. Taken together, we estimate the total atomic displacement of an approximately 15 nm thick subsurface of MAPbI3, where over 90 % of the damage is mainly concentrated within a 5 nm subsurface. For comparison, we simulate the beam damage in MAPbI3 under typical Ga$^+$ FIB processes (30 kV irradiation at an incident angle of 52°). The damage cascade is displayed in an elongated pear shape with the central axis along the trajectory of the primary Ar ions (**Figure 6**). The estimated atomic displacement ranges over 50 nm in the depth of MAPbI$_3$, suggesting that post-polishing processes are necessary before high-resolution microscopy measurements. The contour plots in **Figure 6(c, d)** summarize the distribution profiles for both cases.

The simulation results provide useful insight into beam damage profiles in metal-halide perovskites that can be correlated to the observed subsurface characteristics. For instance, the PL intensities after Ar$^+$ milling drops by about 40 % compared to the unexposed areas (**Figure 2**). Considering the spontaneous photoluminescence to be proportional to the absorption profile[38], we estimate an approximately 40 % of the total PL emission is attributed to the 15 nm-thick surfaces of MAPbI$_3$ under a 244 nm laser beam illumination (**Figure S1**). It appears that the Ar$^+$ milling introduces the atomic displacement (e.g., local disorder, point defects), resulting in an optically inactive "dead layer". On the other hand, TRIM calculates an approximately 47 % of the total energy loss is transferred to the phonon energy in the subsurface. This immense energy deposition in the subsurface of MAPbI$_3$ having relatively low thermal conductivity ($\approx$ 0.3 W/m·K) [39] can substantially increase the local temperature above 200 °C [40-42]. This agrees with the XPS experimental results, where the signature of the presence of PbI$_2$ and its complexes are apparent in the polished sample (**Figure 6b**).

3. SUMMARY AND CONCLUSIONS

In summary, we examine the effects of ion beam exposure on metal-halide perovskite by measuring the subsurface characteristics. We find that the irradiation lowers the bandgap PL emission in $MAPbI_3$ by 40 % while increasing the surface potential. Several signatures of the chemical modification are observed at the $MAPbI_3$ surface. XPS analysis shows the organic components (e.g., C, N) almost vanish while enhancing inorganic signatures after the beam exposure, leading to a lead-rich, iodine-deficient surface. Our experiments, together with SRIM/TRIM simulations, suggest that the presence of $PbI_2$ and the complexes on the subsurface is associated with the "dead layer" introduced by the ion beam via atomic displacement and local heating. Specifically, an approximately 15 nm thick subsurface of $MAPbI_3$ can be damaged by a 4 kV $Ar^+$ beam irradiated at an incident angle of 3º. Our findings convey that the practical use of ion beams requires care for perovskite measurements under accelerated stressors. The thickness of the "dead layer" can be reduced by mitigating the local heating (e.g., cryogenic sample stage) and the kinetic beam energy (e.g., low beam voltage). Despite the subsurface modification, our work supports the rest of the $MAPbI_3$ bulk retains the properties, as observed in PL. This implies that FIB-prepared perovskite samples could serve in studying the properties of bulk or buried junctions far away from the altered subsurface. Such surface-insensitive measurement approaches include two-photon PL and micro LBIC (laser beam-induced current), where the wavelengths of the laser beams can tune the profiles of the photon absorption and excess carrier generations.

## 4. EXPERIMENTAL SECTION

***PSC fabrication***: Glass substrates coated with fluorine-doped tin oxide (FTO; a thickness of 200 nm) were cleaned with acetone, isopropyl alcohol (IPA), and deionized (DI) water. The $N_2$-dried substrates were further cleaned with oxygen ($O_2$) plasma. A 35 nm $SnO_2$ (ETL) was deposited on FTO by the atomic layer deposition (ALD) system (Picosun) using high purity (99.9999 %) tetrakis(dimethylamino)tin(iv) and deionized water as precursors. $MAPbI_3$ precursor ink was synthesized by mixing 1 mM lead iodide ($PbI_2$) and 1 mM methylammonium iodide (MAI) in Dimethylformamide (DMF) with an additive of 0.02 mM lead bisdicyanoamide ($Pb(SCN)_2$) in dimethyl sulfoxide (DMSO). This $MAPbI_3$ ink was spun coated on ETL at a spin-speed of 500 rpm for 3 s, followed by 4,000 rpm for 50 s. During the second spin step (i.e., about 11 s from the first spinning), anti-solvent diethyl ether (720 µL) was dropped onto the center of the sample [43]. The sample was cured on a hotplate at 60 °C for 3 minutes, followed by 100 °C for 10 minutes. The samples were stored under $N_2$ environmental conditions in dark before $Ar^+$ polishing and characterizations.

***Shallow angle Ar ion beam polishing***: We performed a series of broad $Ar^+$ milling (Fischione Model 1060) to reduce the surface roughness of the $MAPbI_3$ layer. This system is equipped with two ion sources, where the angle can be adjusted from 0° to 10° to the direction parallel to the sample surface. We used a single ion gun irradiating at 3° without rotation in this experiment. High-purity (99.999 %) argon gas was used for the ion sources at a base pressure of $3.8 \times 10^{-4}$ Torr. The beam current was set to ≈ 50 µA for all samples. Since the Faraday cage measuring the current was located far beyond the sample stage, we used the magnitude of the current only for checking the proper system operation, not for quantitative estimation of the beam energy delivered to the specimen. MAPbI3 samples were polished by Ar$^+$ beams (4 kV irradiating at 3°) for 10 minutes at room temperature.

***Hyperspectroscopy photoluminescence (PL) microscopy***: We used a 224 nm (5.08 eV) UV light source that produced a nominal power of approximately ten µW on the sample (LEXEL Quantum 8 SHG CW deep UV and tunable visible argon ion laser, Cambridge Laser Laboratories Inc.). The PL emission was collected using a 600 mm$^{-1}$ holographic grating with a 40× (numerical aperture [N. A.] = 0.5) reflective objective lens (LMM40X-UVV, ThorLabs). The confocal hole diameter was set to 300 mm. Point PL spectra were taken by rastering over a 50 mm diameter area using fast scanning mirrors as the mean of 2 spectra recorded for 0.5 seconds each. PL mapping was performed at a pixel size of 333 mm in the x and y-directions. The hyperspectroscopy PL signals were collected on a 50 mm diameter area of each spatial pixel. The focal height was obtained at each real space pixel by optimizing the PL signal in the spectral window (760 nm ~ 770 nm) over a 150 mm height range with 0.25 mm z-axis resolution.

***Kelvin-probe force microscopy***: KPFM was conducted using Bruker PFQNE. The work function values of the PSC samples were calculated using the surface potentials of gold (5.1 eV) and highly oriented pyrolytic graphite (HOPG; 4.68 eV). Sample maps were collected with a feedback force setpoint of 8 nN with a scan rate of 0.5 Hz.

***X-ray photoemission spectroscopy***: XPS measurements were carried out using an Al Kα X-ray source (1.486 KeV; Kratos Axis Ultra DLD). The sample was first placed in a load-lock and pumped down to a vacuum pressure of $1 \times 10^{-7}$ Torr and transferred to the analysis chamber (< $6 \times 10^{-10}$ Torr). The emission current was 8 mA, and the anode voltage was 15 kV. A typical beam spot size for the XPS was 300 µm × 700 µm. A neutral gun was employed to eliminate the charge effect during the XPS measurement. The XPS signals were acquired with an energy resolution of 1 eV for survey scans and 0.1 eV for high-resolution spectra. The measured spectra were calibrated using the core level peak of C 1s (C-C bonds) at 284.5 eV. Peak fitting and quantitative analysis (e.g., atomic concentration of each element) were performed using a commercial software with the average matrix relative sensitivity factors (CasaXPS, Origin Pro).

***Thermal degradation***: Qualitative thermal degradation of PSC of $Ar^+$ polished samples was examined using a heat lamp. We placed a ceramic lamp heater (150 W; a diameter of 7 cm) above the sample stage equipped with a K-type thermal couple probe. The distance between the top of the sample and the bottom of the lamp was approximately 9 cm, providing relatively uniform heat distribution in the sample stage. When the stage temperature reached the set value (e.g., 150 °C), we located $Ar^+$ polished and control samples in proximity under the lamp. A series of pictures were taken at a few minute intervals for an hour until the samples were fully degraded and turned color from dark brown to yellow (e.g., $PbI_2$).

***Monte Carlo simulations***: We performed Monte Carlo simulations using a software package to study the $Ar^+$ beam interaction with metal-halide perovskites (SRIM/TRIM: Stopping and Range of Ions in Matter / Transport and Range of Ions in Matter). SRIM/TRIM covers multiple inelastic collisions, and it models range, damage, recoils, thermal effects, and sputtering. The standard TRIM database was used to set the simulation configuration. The injected ion data was set to Ar (atomic number of 18, atomic mass of 39.962u), whereas a 500 nm-thick $MAPbI_3$ was set to the target material (density of 4.16 $g/cm^3$) [37]. The standard values of the atomic number, atomic weight, displacement energy, lattice energy, and surface energy for each component of $MAPbI_3$ were followed in the TRIM library without modification. Our model used 200,000 ions for each simulation. The extracted 2D/3D raw datasets were analyzed using MATLAB.

**Figure 1**.

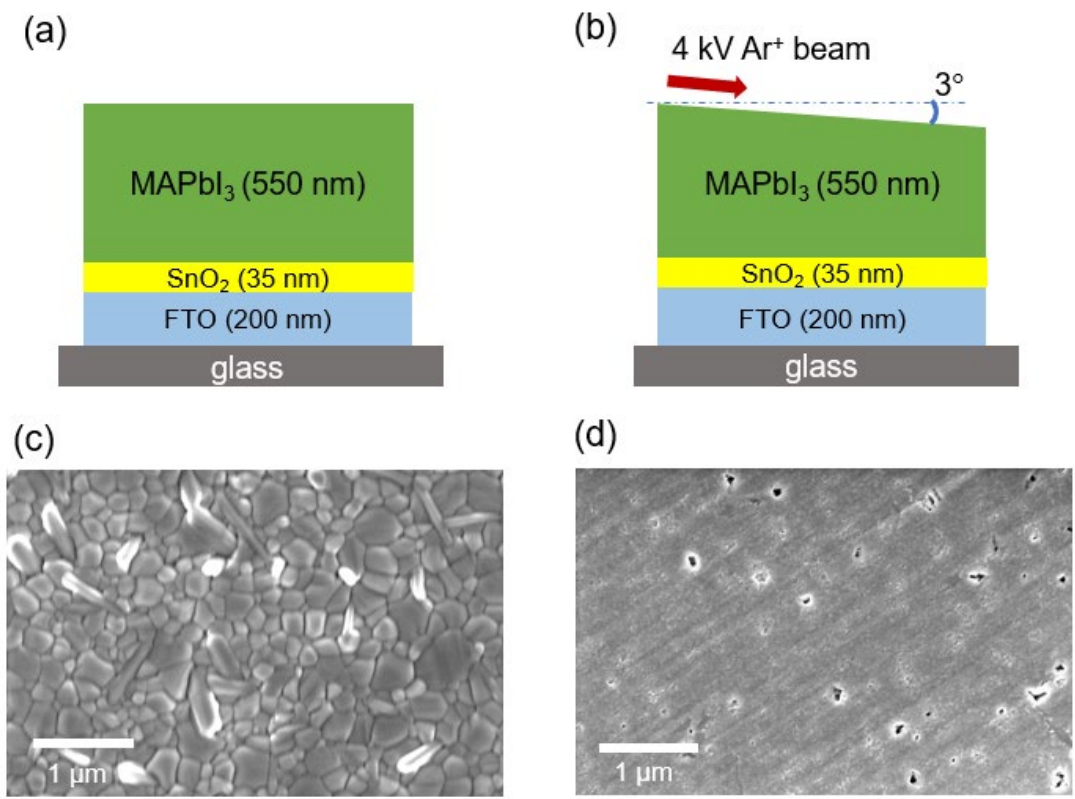

Schematics of methylammonium lead iodide (MAPbI$_3$) samples: (a) pristine and (b) polished using Ar$^+$ beams. Representative scanning electron microscopy (SEM) images of MAPbI$_3$ (c) pristine and (d) polished samples.

**Figure 2.**

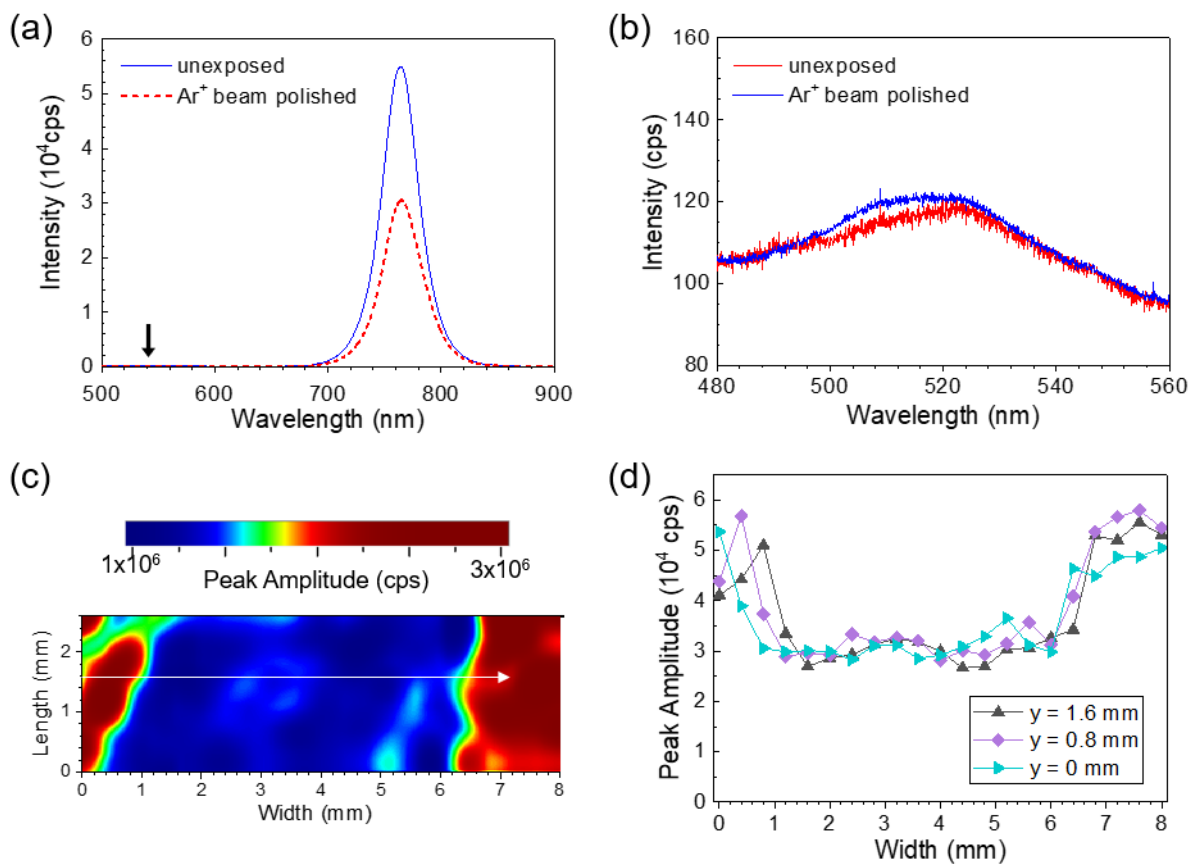

(a) Photoluminescence (PL) responses of pristine and polished samples. The prominent peak 765 nm (1.62 eV) is associated to the bandgap emission. The PL region near 500 nm (marked with an arrow in black) is replotted in (b). The wavelength 516 nm corresponds to 2.4 eV bandgap of $PbI_2$. (c) PL map plotted with the peak intensities at 765 nm. The center area (in blue) was exposed to $Ar^+$ beam, showing a lower PL intensity. The other regions (in red) were not exposed to the ion beam. (d) Intensity profiles along the width of the sample at the length of 0 mm, 8 mm, and 1.6 mm (see white arrow in (c)).

**Figure 3.**

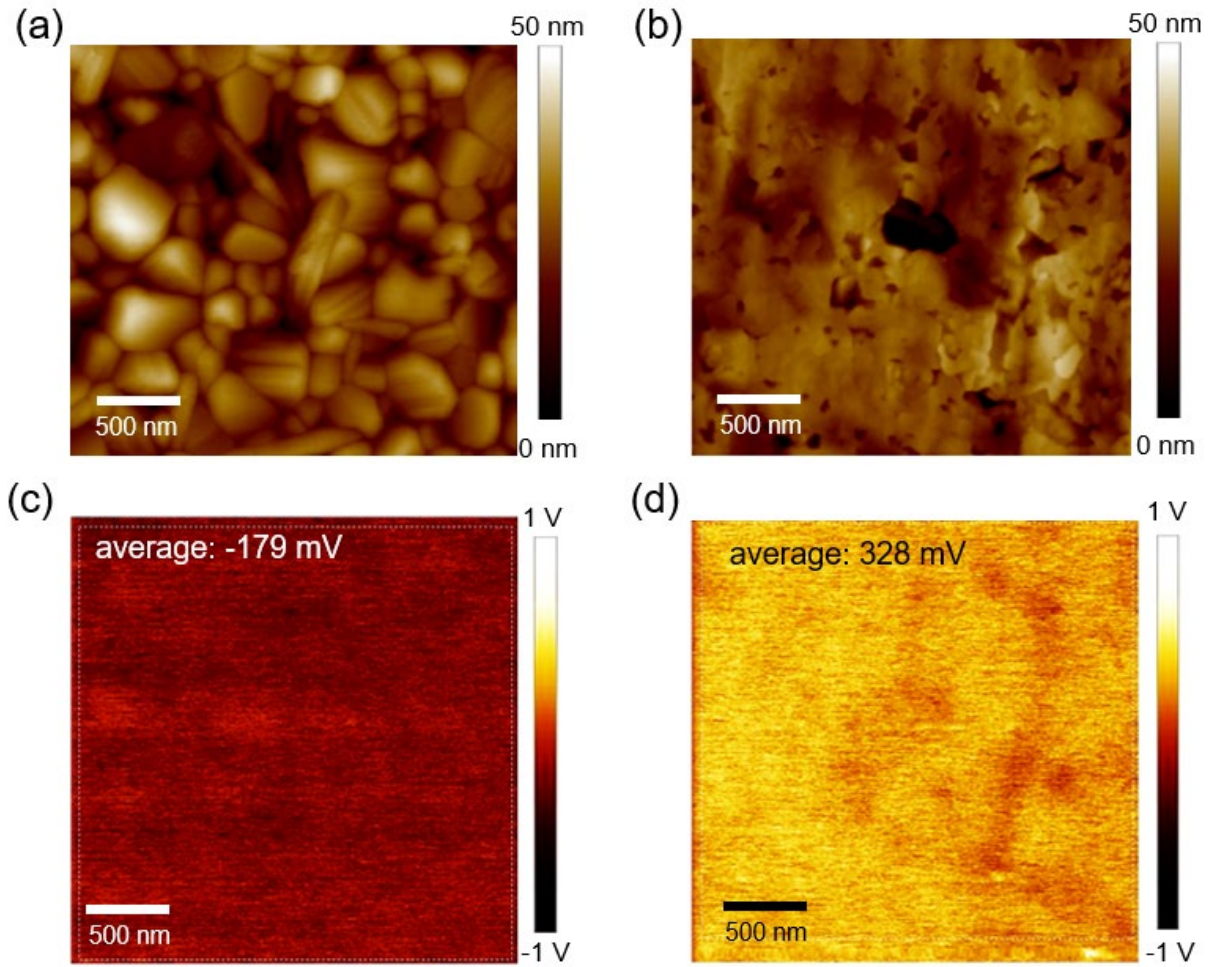

Topography of MAPbI$_3$ obtained using atomic force microscopy (AFM) of (a) pristine and (b) polished samples. The peak-to-valley surface roughness reduces from 100 nm to 10 nm with the polishing processes. (c) and (d) shows the contact potential difference (CPD) simultaneously collected with the AFM images in (a) and (b). The average CPD of the pristine surface (-179 mV ± 150 mV) increases to 328 mV ± 150 mV after polishing.

**Figure 4.**

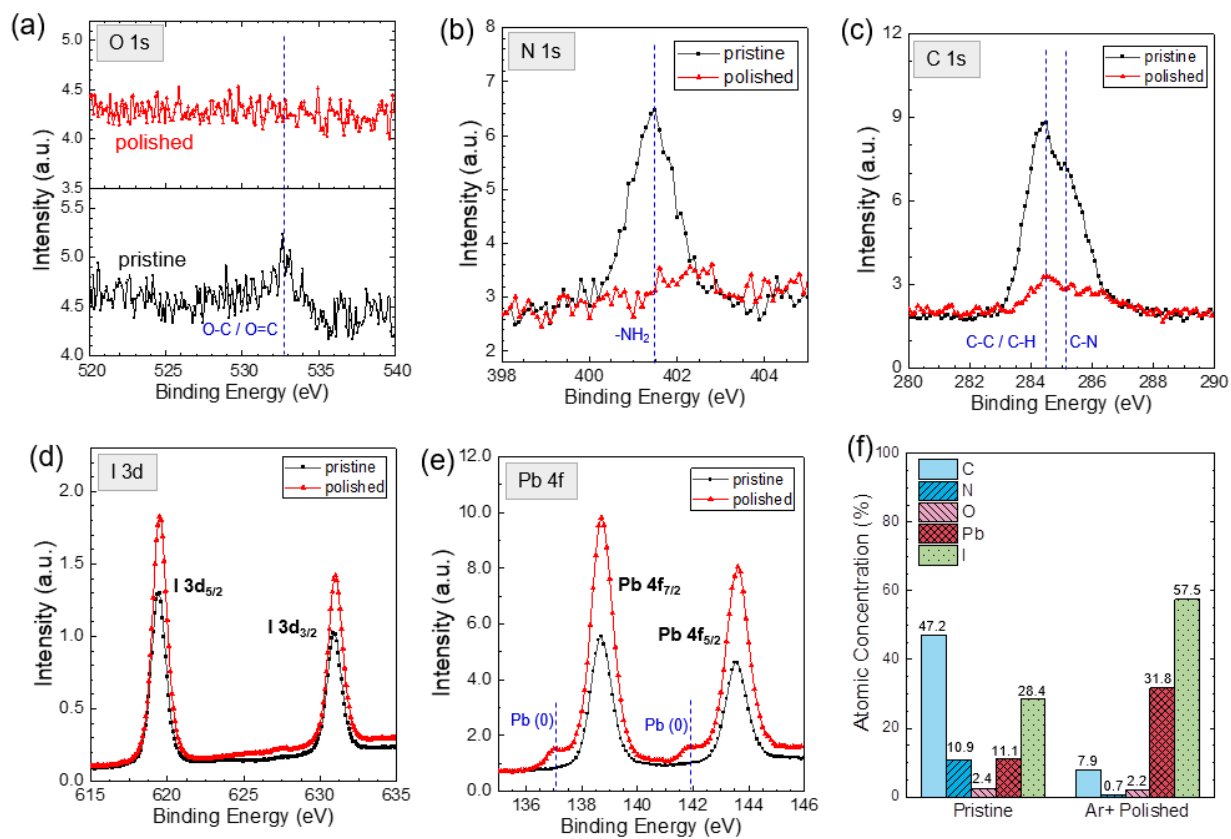

High-resolution XPS (x-ray photoemission spectroscopy) profiles of pristine (black dots) and polished (red squares) samples. Significant decreases of the organic compositions are seen with the Ar+ polished sample: (a) O 1s, (b) N 1s, and (c) C 1s. The XPS signatures of (d) iodine I $3d_{3/2}$ and I $3d_{5/2}$) and (f) lead (Pb $4f_{5/2}$ and Pb $4f_{7/2}$) increase with the polishing. The additional peaks near the shoulders of the Pb 4f doublet peaks infer the introduction of metallic lead [Pb (0); 137.1 eV and 142 eV] and oxidized lead [Pb (II); 138.7 eV and 143.6 eV] under Ar$^+$ beam irradiation. (f) Summary of atomic percentage of MAPbI$_3$ "before (pristine)" and "after (polished)" the ion beam irradiation.

**Figure 5.**

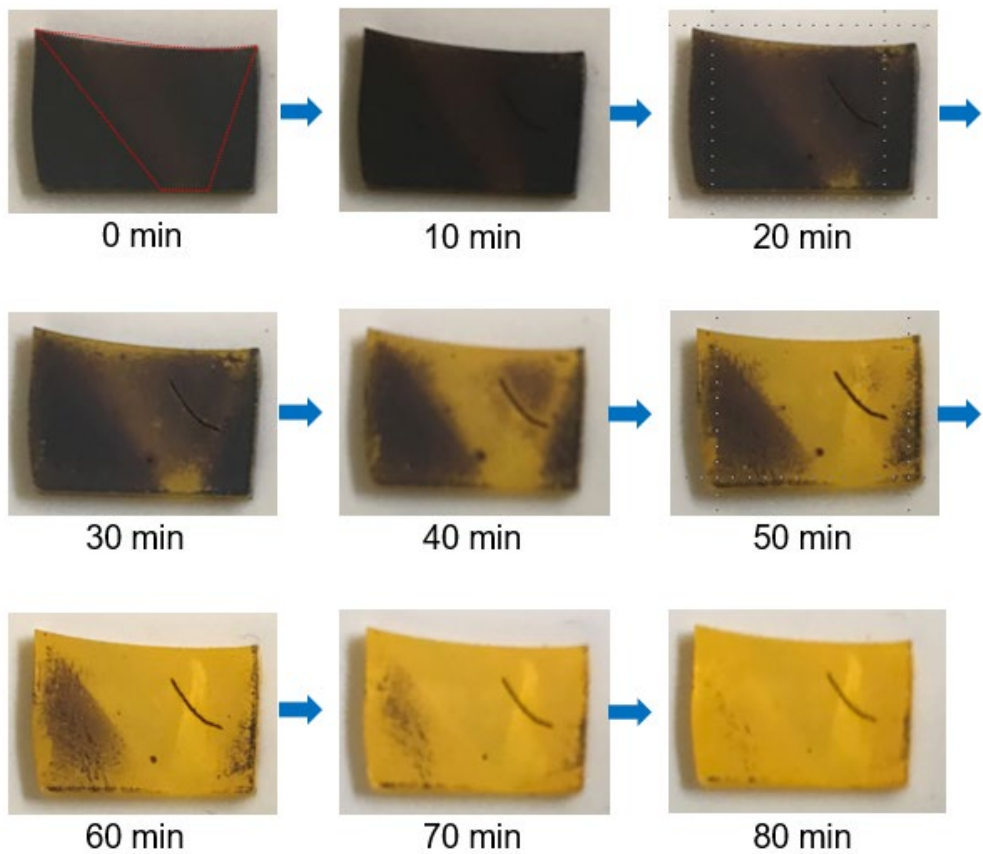

A series of photographs of MAPbI$_3$ taken during thermal aging processes. The polished region is seen in the trapezoidal region of the sample (red dot lines).

**Figure 6.**

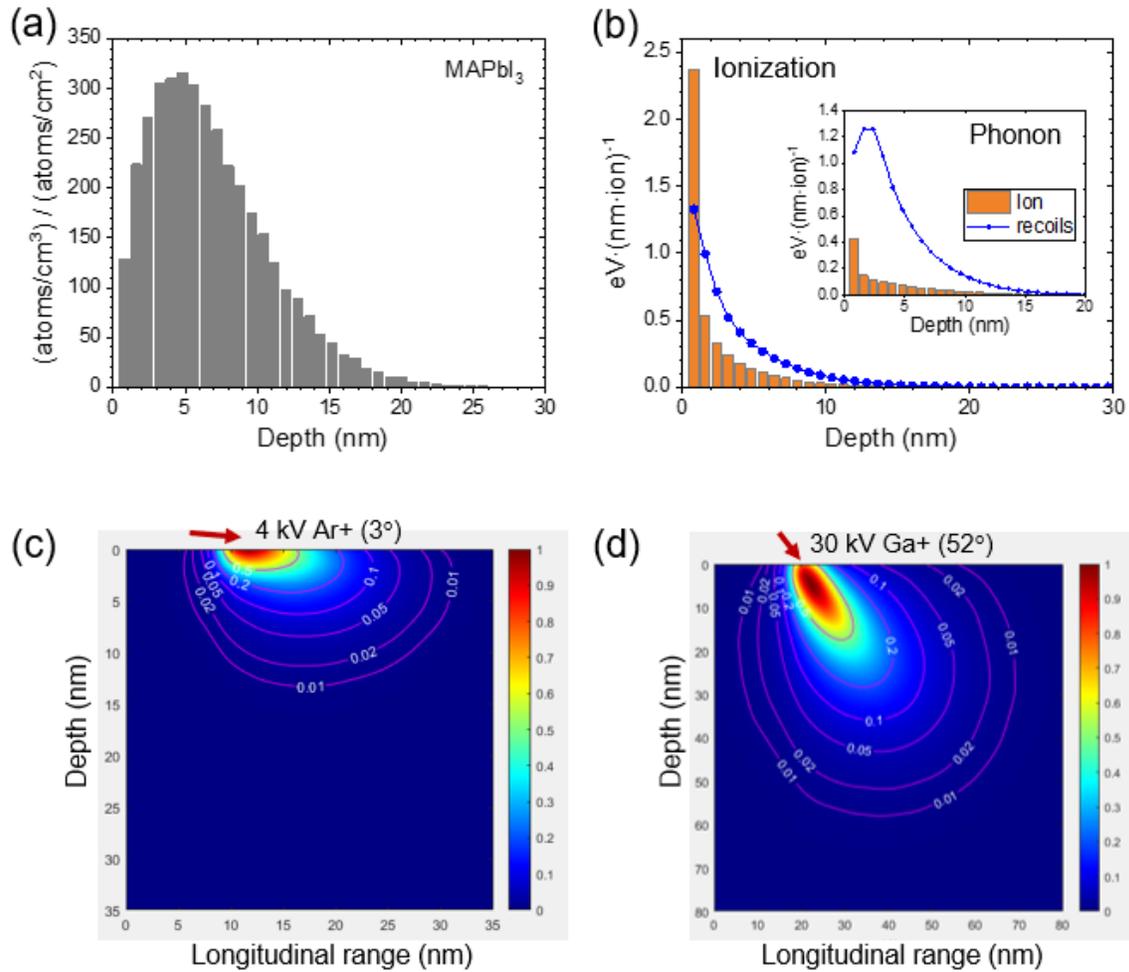

SRIM/TRIM simulations estimate the ion beam damage in MAPbI$_3$. (a) Distribution of total ion range in MAPbI$_3$. (b) Plots showing the ionization profiles of the primary Ar$^+$ ions and their recoils. The inset displays the energy loss of the primary ions and the recoils to phonons. (c) Contour plots showing the distribution of total atomic displacement by (c) 4 kV Ar$^+$ ion at 3º and (d) 30 kV Ga$^+$ ion at 52° irradiated onto the MAPbI$_3$ surface. The simulations predict a damage layer of ≈ 15 nm for the Ar$^+$ beam and ≈ 60 nm for the Ga$^+$ beam.

## ASSOCIATED CONTENT

**Supporting Information**.

FDTD simulation and XPS survey (PDF)

## AUTHOR INFORMATION

**Corresponding Author**

Heayoung P. Yoon - Electrical and Computer Engineering & Materials Science and Engineering, University of Utah, Salt Lake City, UT 84112, USA, Email: heayoung.yoon@utah.edu

**Present Addresses**

Yu-Lin Hsu - Department of Electrical and Computer Engineering, University of Utah, Salt Lake City, UT 84112, USA

Chongwen Li - Department of Physics and Astronomy and Wright Center for Photovoltaics Innovation and Commercialization, The University of Toledo, Toledo, Ohio 43606, USA

Andrew C. Jones - Center for Integrated Nanotechnologies, Los Alamos National Laboratory, Los Alamos, NM 87545, USA

Michael T. Pratt - Center for Integrated Nanotechnologies, Los Alamos National Laboratory, Los Alamos, NM 87545, USA

Ashif Chowdhury - Electrical and Computer Engineering, University of Utah, Salt Lake City, UT 84112, USA

Yanfa Yan: Department of Physics and Astronomy and Wright Center for Photovoltaics Innovation and Commercialization, The University of Toledo, Toledo, Ohio 43606, USA

ACKNOWLEDGMENT

The authors thank K. Powell, P. Perez, and P. Lysak for valuable discussions and assisting analysis during this work. This research was supported by the U.S. Department of Energy's Office of Energy Efficiency and Renewable Energy (EERE) under the DE-FOA-0002064 program award number DE-EE0008985. This work was performed, in part, at the Center for Integrated Nanotechnologies, an Office of Science User Facility operated for the U.S. Department of Energy (DOE) Office of Science. Los Alamos National Laboratory, an affirmative action equal opportunity employer, is managed by Triad National Security, LLC for the U.S. Department of Energy's NNSA, under contract 89233218CNA000001. M. Prett. acknowledges support from Laboratory Directed Research and Development (LDRD) award 20210640ECR. Y. Yan. acknowledges support of NSF under Award No. DMR-1807818. This work was support by the USTAR shared facilities at the University of Utah, in part, by the MRSEC Program of NSF under Award No. DMR-1121252. H. Yoon acknowledges the support from the NSF CAREER Award No. 2048152.

REFERENCES

1. Giannuzzi, L. A.; Stevie, F. A., A review of focused ion beam milling techniques for TEM specimen preparation. *Micron* **1999,** *30* (3), 197-204.

2. Huang, Z., Combining Ar ion milling with FIB lift-out techniques to prepare high quality site-specific TEM samples. *Journal of Microscopy* **2004,** *215* (3), 219-223.

3. Kim, C.-S.; Ahn, S.-H.; Jang, D.-Y., Review: Developments in micro/nanoscale fabrication by focused ion beams. *Vacuum* **2012,** *86* (8), 1014-1035.

4. Giannuzzi, L. A.; Stevie, F. A., *Introduction to focused ion beams : instrumentation, theory, techniques, and practice*. Springer: New York, 2005.

5. Bischoff, L.; Mazarov, P.; Bruchhaus, L.; Gierak, J., Liquid metal alloy ion sources—An alternative for focussed ion beam technology. *Applied Physics Reviews* **2016,** *3* (2), 021101.

6. Kato, N. I., Reducing focused ion beam damage to transmission electron microscopy samples. *Journal of Electron Microscopy* **2004,** *53* (5), 451-458.

7. Turner, E. M.; Sapkota, K. R.; Hatem, C.; Lu, P.; Wang, G. T.; Jones, K. S., Wet-chemical etching of FIB lift-out TEM lamellae for damage-free analysis of 3-D nanostructures. *Ultramicroscopy* **2020,** *216*, 113049.

8. Campin, M. J.; Bonifacio, C. S.; Nowakowski, P.; Fischione, P. E.; Giannuzzi, L. A. In *Narrow-Beam Argon Ion Milling of Ex Situ Lift-Out FIB Specimens Mounted on Various Carbon-Supported Grids*, 2018; ASM International: 2018; pp 339-344.

9. Meng, X.; Tian, X.; Zhang, S.; Zhou, J.; Zhang, Y.; Liu, Z.; Chen, W., In Situ Characterization for Understanding the Degradation in Perovskite Solar Cells. *Solar RRL* **2022,** *6* (7), 2200280.

10. Ran, J.; Dyck, O.; Wang, X.; Yang, B.; Geohegan, D. B.; Xiao, K., Electron-Beam-Related Studies of Halide Perovskites: Challenges and Opportunities. *Advanced Energy Materials* **2020,** *10* (26), 1903191.

11. Di Girolamo, D.; Phung, N.; Kosasih, F. U.; Di Giacomo, F.; Matteocci, F.; Smith, J. A.; Flatken, M. A.; Köbler, H.; Turren Cruz, S. H.; Mattoni, A.; Cinà, L.; Rech, B.; Latini, A.; Divitini, G.; Ducati, C.; Di Carlo, A.; Dini, D.; Abate, A., Ion Migration-Induced Amorphization and Phase Segregation as a Degradation Mechanism in Planar Perovskite Solar Cells. *Advanced Energy Materials* **2020,** *10* (25), 2000310.

12. Seo, Y.-H.; Kim, J. H.; Kim, D.-H.; Chung, H.-S.; Na, S.-I., In situ TEM observation of the heat–induced degradation of single– and triple–cation planar perovskite solar cells. *Nano Energy* **2020,** *77*, 105164.

13.	Byeon, J.; Kim, J.; Kim, J.-Y.; Lee, G.; Bang, K.; Ahn, N.; Choi, M., Charge transport layer-dependent electronic band bending in perovskite solar cells and its correlation to light-induced device degradation. *ACS Energy Letters* **2020,** *5* (8), 2580-2589.

14.	Zhao, Y.; Zhou, W.; Han, Z.; Yu, D.; Zhao, Q., Effects of ion migration and improvement strategies for the operational stability of perovskite solar cells. *Physical Chemistry Chemical Physics* **2021,** *23* (1), 94-106.

15.	Chen, S.; Zhang, Y.; Zhao, J.; Mi, Z.; Zhang, J.; Cao, J.; Feng, J.; Zhang, G.; Qi, J.; Li, J.; Gao, P., Transmission electron microscopy of organic-inorganic hybrid perovskites: myths and truths. *Science Bulletin* **2020,** *65* (19), 1643-1649.

16.	Deng, Y.-H., Perovskite decomposition and missing crystal planes in HRTEM. *Nature* **2021,** *594* (7862), E6-E7.

17.	Kosasih, F. U.; Divitini, G.; Orri, J. F.; Tennyson, E. M.; Kusch, G.; Oliver, R. A.; Stranks, S. D.; Ducati, C., Optical emission from focused ion beam milled halide perovskite device cross-sections. *Microscopy Research and Technique* **2022,** *85* (6), 2351-2355.

18.	Zhou, Y.; Yang, M.; Vasiliev, A. L.; Garces, H. F.; Zhao, Y.; Wang, D.; Pang, S.; Zhu, K.; Padture, N. P., Growth control of compact CH 3 NH 3 PbI 3 thin films via enhanced solid-state precursor reaction for efficient planar perovskite solar cells. *Journal of Materials Chemistry A* **2015,** *3* (17), 9249-9256.

19.	Ghimire, K.; Zhao, D.; Cimaroli, A.; Ke, W.; Yan, Y.; Podraza, N. J., Optical monitoring of CH3NH3PbI3 thin films upon atmospheric exposure. *Journal of Physics D: Applied Physics* **2016,** *49* (40), 405102.

20.	https://www.lumerical.com/, "Lumerical computation solutions: Tcad software," ed.

21.	Shirayama, M.; Kadowaki, H.; Miyadera, T.; Sugita, T.; Tamakoshi, M.; Kato, M.; Fujiseki, T.; Murata, D.; Hara, S.; Murakami, T. N.; Fujimoto, S.; Chikamatsu, M.; Fujiwara, H., Optical Transitions in Hybrid Perovskite Solar Cells: Ellipsometry, Density Functional Theory, and Quantum Efficiency Analyses for ${\mathrm{CH}}_{3}{\mathrm{NH}}_{3}{\mathrm{PbI}}_{3}$. *Physical Review Applied* **2016,** *5* (1), 014012.

22.	Alvarado-Leaños, A. L.; Cortecchia, D.; Folpini, G.; Srimath Kandada, A. R.; Petrozza, A., Optical Gain of Lead Halide Perovskites Measured via the Variable Stripe Length Method: What We Can Learn and How to Avoid Pitfalls. *Advanced Optical Materials* **2021,** *9* (18), 2001773.

23.	Chen, X.; Xia, Y.; Huang, Q.; Li, Z.; Mei, A.; Hu, Y.; Wang, T.; Cheacharoen, R.; Rong, Y.; Han, H., Tailoring the Dimensionality of Hybrid Perovskites in Mesoporous Carbon Electrodes for Type-II Band Alignment and Enhanced Performance of Printable Hole-Conductor-Free Perovskite Solar Cells. *Advanced Energy Materials* **2021,** *11* (18), 2100292.

24. Chandra Shakher, P., Application of Atomic Force Microscopy in Organic and Perovskite Photovoltaics. In *Recent Developments in Atomic Force Microscopy and Raman Spectroscopy for Materials Characterization*, Chandra Shakher, P.; Samir, K., Eds. IntechOpen: Rijeka, 2021; p Ch. 1.

25. Lanzoni, E. M.; Gallet, T.; Spindler, C.; Ramirez, O.; Boumenou, C. K.; Siebentritt, S.; Redinger, A., The impact of Kelvin probe force microscopy operation modes and environment on grain boundary band bending in perovskite and Cu (In, Ga) Se2 solar cells. *Nano Energy* **2021,** *88*, 106270.

26. Panchal, V.; Pearce, R.; Yakimova, R.; Tzalenchuk, A.; Kazakova, O., Standardization of surface potential measurements of graphene domains. *Scientific Reports* **2013,** *3* (1), 2597.

27. Gallet, T.; Lanzoni, E. M.; Redinger, A. In *Effects of Annealing and Light on Co-evaporated Methylammonium Lead Iodide Perovskites using Kelvin Probe Force Microscopy in Ultra-High Vacuum*, 2019 IEEE 46th Photovoltaic Specialists Conference (PVSC), 2019/06/16/21; 2019; pp 1477-1482.

28. Olthof, S.; Meerholz, K., Substrate-dependent electronic structure and film formation of MAPbI3 perovskites. *Scientific Reports* **2017,** *7* (1), 40267.

29. Guo, X.; McCleese, C.; Kolodziej, C.; Samia, A. C. S.; Zhao, Y.; Burda, C., Identification and characterization of the intermediate phase in hybrid organic–inorganic MAPbI3 perovskite. *Dalton Transactions* **2016,** *45* (9), 3806-3813.

30. Rajendra Kumar, G.; Dennyson Savariraj, A.; Karthick, S. N.; Selvam, S.; Balamuralitharan, B.; Kim, H.-J.; Viswanathan, K. K.; Vijaykumar, M.; Prabakar, K., Phase transition kinetics and surface binding states of methylammonium lead iodide perovskite. *Physical Chemistry Chemical Physics* **2016,** *18* (10), 7284-7292.

31. Lin, W.-C.; Lo, W.-C.; Li, J.-X.; Wang, Y.-K.; Tang, J.-F.; Fong, Z.-Y., In situ XPS investigation of the X-ray-triggered decomposition of perovskites in ultrahigh vacuum condition. *npj Materials Degradation* **2021,** *5* (1), 13.

32. Kim, T. W.; Shibayama, N.; Cojocaru, L.; Uchida, S.; Kondo, T.; Segawa, H., Real-Time In Situ Observation of Microstructural Change in Organometal Halide Perovskite Induced by Thermal Degradation. *Advanced Functional Materials* **2018,** *28* (42), 1804039.

33. Kim, N.-K.; Min, Y. H.; Noh, S.; Cho, E.; Jeong, G.; Joo, M.; Ahn, S.-W.; Lee, J. S.; Kim, S.; Ihm, K.; Ahn, H.; Kang, Y.; Lee, H.-S.; Kim, D., Investigation of Thermally Induced Degradation in CH3NH3PbI3 Perovskite Solar Cells using In-situ Synchrotron Radiation Analysis. *Scientific Reports* **2017,** *7* (1), 4645.

34. Dunfield, S. P.; Bliss, L.; Zhang, F.; Luther, J. M.; Zhu, K.; van Hest, M. F. A. M.; Reese, M. O.; Berry, J. J., From Defects to Degradation: A Mechanistic Understanding of Degradation in Perovskite Solar Cell Devices and Modules. *Advanced Energy Materials* **2020,** *10* (26), 1904054.

35. Ziegler, J. F.; Ziegler, M. D.; Biersack, J. P., SRIM – The stopping and range of ions in matter (2010). *19th International Conference on Ion Beam Analysis* **2010,** *268* (11), 1818-1823.

36. Luo, P.; Sun, X.-Y.; Li, Y.; Yang, L.; Shao, W.-Z.; Zhen, L.; Xu, C.-Y., Correlation between Structural Evolution and Device Performance of CH3NH3PbI3 Solar Cells under Proton Irradiation. *ACS Applied Energy Materials* **2021,** *4* (12), 13504-13515.

37. Stuckelberger, M.; Nietzold, T.; Hall, G. N.; West, B.; Werner, J.; Niesen, B.; Ballif, C.; Rose, V.; Fenning, D. P.; Bertoni, M. I. In *Elemental distribution and charge collection at the nanoscale on perovskite solar cells*, 2016 IEEE 43rd Photovoltaic Specialists Conference (PVSC), 2016/06/05/10; 2016; pp 1191-1196.

38. van Roosbroeck, W.; Shockley, W., Photon-Radiative Recombination of Electrons and Holes in Germanium. *Physical Review* **1954,** *94* (6), 1558-1560.

39. Heiderhoff, R.; Haeger, T.; Pourdavoud, N.; Hu, T.; Al-Khafaji, M.; Mayer, A.; Chen, Y.; Scheer, H.-C.; Riedl, T., Thermal Conductivity of Methylammonium Lead Halide Perovskite Single Crystals and Thin Films: A Comparative Study. *The Journal of Physical Chemistry C* **2017,** *121* (51), 28306-28311.

40. Süess, M. J.; Mueller, E.; Wepf, R., Minimization of amorphous layer in Ar+ ion milling for UHR-EM. *Ultramicroscopy* **2011,** *111* (8), 1224-1232.

41. Barber, D. J., Radiation damage in ion-milled specimens: characteristics, effects and methods of damage limitation. *Ultramicroscopy* **1993,** *52* (1), 101-125.

42. Park, Y. M.; Ko, D.-S.; Yi, K.-W.; Petrov, I.; Kim, Y.-W., Measurement and estimation of temperature rise in TEM sample during ion milling. *Ultramicroscopy* **2007,** *107* (8), 663-668.

43. Wang, C.; Xiao, C.; Yu, Y.; Zhao, D.; Awni, R. A.; Grice, C. R.; Ghimire, K.; Constantinou, I.; Liao, W.; Cimaroli, A. J.; Liu, P.; Chen, J.; Podraza, N. J.; Jiang, C.-S.; Al-Jassim, M. M.; Zhao, X.; Yan, Y., Understanding and Eliminating Hysteresis for Highly Efficient Planar Perovskite Solar Cells. *Advanced Energy Materials* **2017,** *7* (17), 1700414.